\documentclass[aps,showpacs,twocolumn,floatfix,epsf]{revtex4} 
\usepackage{graphicx}        

\begin{document}
   
 
\title{Josephson oscillation and induced collapse in an attractive
Bose-Einstein condensate}

\author{Sadhan K. Adhikari\thanks{e-mail: adhikari@ift.unesp.br}}
\affiliation{Instituto de F\'{\i}sica Te\'orica, Universidade Estadual
Paulista, 01.405-900 S\~ao Paulo, S\~ao Paulo, Brazil}

\date{\today}

\begin{abstract}


{ Using the axially-symmetric time-dependent Gross-Pitaevskii
equation we study the Josephson oscillation of an attractive Bose-Einstein
condensate (BEC)  in a one-dimensional periodic optical-lattice
potential. 
We find that the  Josephson  frequency is
virtually independent of the
number of atoms in the BEC and of the inter-atomic interaction (attractive
or repulsive). 
We study the dependence of  Josephson frequency on the laser wave
length  and the
strength of the optical-lattice potential. For a fixed laser wave length
(795 nm),
the  Josephson frequency decreases with increasing strength as found in
the experiment of Cataliotti {\it et al.}  [Science {\bf
293}, 843 (2001)]. 
For a fixed strength, the  Josephson frequency remains
essentially unchanged for a reasonable  variation of 
laser wave length around 800 nm. 
However,  the Josephson oscillation is disrupted with  the increase of
laser wave length  beyond 2000 nm leading to a collapse of a
sufficiently attractive 
BEC. These features of  Josephson oscillation can be tested experimentally
with present set ups.
} 

 
\end{abstract}

\maketitle

\section{Introduction}
 
The observation of an oscillating Josephson current across
 an one-dimensional array of potential wells, usually generated by a
standing-wave laser field and commonly known as an optical-lattice
potential, in a trapped cigar-shaped Bose-Einstein condensate (BEC) by
Cataliotti {\it et al.} \cite{cata} was the first manifestation of this
phenomenon in trapped neutral bosons.  Until then the Josephson effect
was confirmed in superconductors with charged electrons and in liquid
helium \cite{3}.

In the experiment of Cataliotti {\it et al.} \cite{cata} a Josephson
oscillation was initiated in a repulsive $^{87}$Rb BEC formed in a
one-dimensional optical-lattice plus axially-symmetric harmonic traps by
suddenly displacing the harmonic trap along the axial direction. They
investigated the evolution of the Josephson frequency with the strength of
the
optical-lattice potential and found that, for a fixed laser wave length, 
the Josephson frequency reduces with
the increase of the strength. 

There have been previous theoretical studies of Josephson oscillation to
explain
different features of the  experiment of Cataliotti {\it et al.}
\cite{cata} using the numerical
solution of the time-dependent mean-field Gross-Pitaevskii (GP) equation
\cite{8}.
One \cite{cata,cata2,cata3} and three-dimensional 
\cite{sad1} models
based on
the GP equation have previously
been used to study different aspects of Josephson oscillation. 
There have been other  theoretical studies of  Josephson oscillation 
in a trapped BEC \cite{str,str1,th1} using different  approaches and
under different conditions.

In this paper we address several interesting
features of Josephson oscillation in a trapped BEC as in the
experiment of
Cataliotti {\it et al.} \cite{cata}.
How does the Josephson frequency 
evolve with the number of atoms and
atomic interaction (attractive or repulsive) of the condensate?
How does the Josephson frequency depend on  the 
intensity (strength) and wave length of the standing-wave 
laser beam used to create the periodic
optical-lattice potential? What are the limits on the 
optical-lattice
potential parameters 
for allowing a Josephson oscillation? Is it possible to have a
collapse in an attractive BEC induced by a Josephson
oscillation?

In the actual experiment \cite{cata} and a previous three-dimensional
simulation \cite{sad1} the extraction of  
Josephson frequency  was rather indirect. The BEC in the optical-lattice
potential was allowed to expand and form an interference pattern of two
bright pieces moving in opposite directions from a central piece after a
certain interval of time. The position of the original BEC executing a 
Josephson oscillation in the periodic
optical-lattice  potential  was predicted from the
position of the observed interference pattern. In this indirect fashion
the Josephson oscillation was identified and its frequency
extracted. Although, because of the very small size of the BEC, 
this seems to be the only means for experimental observation, we
employ a direct approach for the study of Josephson oscillation in the
present numerical simulation, which is more accurate and easier to
implement than the indirect approach employed in
Refs. \cite{cata,sad1}. In present  simulation we directly follow
the oscillating condensate without resorting to an expansion.

There have been experimental \cite{don} and theoretical \cite{theory}
studies of the collapse of a
$^{85}$Rb BEC after suddenly turning the atomic interaction from repulsive
to attractive by manipulating an external magnetic field near a Feshbach
resonance \cite{fesh}. There has also been a study of induced collapse of
an
attractive  BEC trapped by a system of laser beams 
in the presence of small
fluctuations of the laser intensity \cite{abdul}. In this paper, to the
best of our knowledge,  
we study
for the first time 
the
possibility of collapse 
of an attractive BEC induced by  Josephson
oscillation. In view of the routine experiments \cite{mo,routine} on BECs
formed in
optical-lattice potentials, it should be possible to verify the
predictions
of the present study on the Josephson oscillation of an  attractive  
BEC.

In Sec.  II  we present the mean-field model based on the
axially-symmetric time-dependent nonlinear 
GP equation. In Sec. III we present the numerical results and finally, in
Sec.  IV we present a summary  of our study.

\section{Mean-field Model}

The time-dependent BEC wave
function $\Psi({\bf r};t)$ at position ${\bf r} $ and time $t $
is described by the following  mean-field nonlinear GP equation
\cite{8}
\begin{eqnarray}\label{a} \left[- i\hbar\frac{\partial
}{\partial t}
-\frac{\hbar^2\nabla^2   }{2m}
+ V({\bf r})
+ gN|\Psi({\bf
r};t)|^2
 \right]\Psi({\bf r};t)=0,
\end{eqnarray}
where $m$
is
the mass and  $N$ the number of atoms in the
condensate,
 $g=4\pi \hbar^2 a/m $ the strength of inter-atomic interaction, with
$a$ the atomic scattering length.  In the presence of the combined
axially-symmetric and periodic
optical-lattice  potentials 
     $  V({\bf
r}) =\frac{1}{2}m \omega ^2(\rho ^2+\nu^2 y^2) +V_{\mbox{opt}}$ where
 $\omega$ is the angular frequency of the harmonic potential 
in the radial direction $\rho$,
$\nu \omega$ that in  the
axial direction $z$, with $\nu$ the aspect ratio, 
and $V_{\mbox{opt}}= s E_R\cos^2 (k_Lz)$ is
the optical-lattice potential  
created with the standing-wave laser field of wavelength 
$\lambda=795$ nm, as in the experiment of Ref. \cite{cata}, 
with $E_R=\hbar^2k_L^2/(2m)$, $k_L=2\pi/\lambda$, and $s$ $ (<12)$
the 
strength. 
The normalization condition  is
$ \int d{\bf r} |\Psi({\bf r};t)|^2 = 1. $

In the axially-symmetric configuration, the wave function
can be written as 
$\Psi({\bf r}, t)= \psi(\rho,z,t)$, where $0\le  \rho < \infty$ is the
radial
variable and $-\infty <z<\infty $ is the axial variable.
Now  transforming to
dimensionless variables $\hat \rho =\sqrt 2 \rho/l$,  
$\hat z=\sqrt 2 z/l$,   $\tau
=t
\omega, $
$l\equiv \sqrt {\hbar/(m\omega)}$,
and
${ \varphi(\hat \rho,\hat z;\tau)} \equiv   
\hat \rho \sqrt{{l^3}/{\sqrt
8}}\psi(\rho,z;t),$  Eq.  (\ref{a}) becomes \cite{9}
\begin{eqnarray}\label{d1}
&\biggr[&-i\frac{\partial
}{\partial \tau} -\frac{\partial^2}{\partial
\hat \rho ^2}+\frac{1}{\hat \rho }\frac{\partial}{\partial \hat \rho}
-\frac{\partial^2}{\partial
\hat z^2}
+\frac{1}{4}\left(\hat \rho ^2+\nu^2 \hat z^2\right)  -{1\over \hat \rho
^2} 
\nonumber \\
&+&
s \frac{E_R}{\hbar\omega}\cos^2(k_Lz)  +
   8\sqrt 2 \pi n\left|\frac {\varphi({\hat \rho ,\hat z};\tau)}{\hat
\rho}\right|^2
 \biggr]\varphi({ \hat \rho,\hat z};\tau)=0, \nonumber \\        
\end{eqnarray}
where the nonlinearity parameter 
$ n=Na /l$. In terms of the 
one-dimensional probability 
 $P(z,t) \equiv 2\pi\- \- \int_0 ^\infty 
d\hat \rho |\varphi(\hat \rho,\hat z,\tau)|^2/\hat \rho $, the
normalization of the
wave 
function 
is given by $\int_{-\infty}^\infty d\hat z P(z,t) = 1.$

The  experiment of Cataliotti {\it et al.} \cite{cata} was performed with 
repulsive $^{87}$Rb atoms in the hyperfine state $F=1,
m_F=-1$. The axial and radial trap frequencies were $\nu \omega =
2\pi \times 9 $ Hz and $ \omega =
2\pi \times 92$ Hz, respectively, with $\nu = 9/92$. In the present
simulation we consider attractive $^{85}$Rb atoms  in the hyperfine state
$F=2, m_F=-2$
with mass quite close
to  $^{87}$Rb atoms. The triplet scattering length $a_T$ for the
$^{85}$Rb-$^{85}$Rb system is negative corresponding to attraction: 
$-500a_0<a_T<-300a_0$ with $a_0=0.5292$ \AA \- the Bohr radius \cite{sca}.
Also, in case of  $^{85}$Rb the atomic interaction
can be modified using a Feshbach resonance \cite{fesh}. So the
$^{85}$Rb-$^{85}$Rb system is the most suitable for studying the
dependence of Josephson oscillation on atomic interaction.

For $^{85}$Rb the
harmonic-oscillator length $l=\sqrt {\hbar/(m\omega)} = 1.14$ $\mu$m and
the
present 
dimensionless length unit  corresponds to $l/\sqrt 2 =0.8$ $\mu$m. The
present
dimensionless time unit corresponds to $\omega ^{-1} =
1/(2\pi\times 92)$ s $=1.73$ ms. Although we perform the calculation in
dimensionless units using Eq. (\ref{d1}), we present the results in
actual physical units using these conversion factors consistent with 
$^{85}$Rb atoms.
In terms of the dimensionless laser wave
length $\lambda _0= \sqrt2\lambda/l \simeq 1$, the dimensionless 
standing-wave energy parameter $E_R/(\hbar \omega)= 4\pi^2/\lambda _0^2$.
Hence in 
dimensionless unit 
the periodic optical-lattice potential  of 
Eq.   (\ref{d1}) is
\begin{equation}\label{pot}
\frac{ V_{\mbox{opt}}}{\hbar
\omega}\equiv s \frac{E_R}{\hbar\omega}\cos^2(k_Lz)
=s \frac{4\pi^2}{\alpha^2\lambda_0^2} 
\left[
\cos^2 \left(
\frac{2\pi}{\alpha \lambda_0}\hat z
\right)
 \right],
\end{equation}
where we have introduced a parameter $\alpha$ to manipulate the laser
wave length. Normally, $\alpha =1$ corresponds to the experimental
situation: $\lambda = 795$ nm. A general value for $\alpha$
simulates a laser wave length of $\alpha \lambda$.

We solve  Eq.  (\ref{d1}) numerically  using a   
split-step time-iteration
method
with  the Crank-Nicholson discretization scheme described recently
\cite{11}.  
We discretize the GP equation typically with time step 0.001
and space step 0.1 spanning $\rho$ from 0 to 7 $\mu$m  and $z$ from $-70$
$\mu$m to
70 $\mu$m, although, sometimes we used smaller steps for obtaining
convergence. 
The time iteration is started with the known harmonic oscillator solution
of  Eq.  (\ref{d1}) for 
 $n=0$: $\varphi(\hat \rho, \hat z) = [\nu
/(8\pi^3)  ]^{1/4}$
$\hat \rho$ $e^{-(\hat \rho^2+\nu \hat z  ^2)/4}$ with chemical potential
$\bar
\mu=(1+\nu /2)$
\cite{9}. For a typical cigar-shaped condensate with $\nu \simeq 0.1$
\cite{cata} $\bar \mu \simeq 1$ is much smaller than the typical depth of
the 
optical-lattice  potential wells $E_R/(\hbar \omega) = 4\pi^2 /\lambda
_0^2 \simeq
40$  so that $\bar \mu
<<
E_R/(\hbar \omega)$ and the  passage of condensate atoms from one well to
other can only
proceed through quantum tunneling. 
The
nonlinearity  $n$ as well as the optical-lattice potential parameter
$s$ 
are  slowly increased by equal amounts in $10000n$ steps of 
time iteration until the desired value of nonlinearity and optical-lattice
potentials are  attained. Then, without changing any
parameter, the solution so obtained is iterated 50 000 times so that a
stable
solution  is obtained 
independent of the initial input
and time and space steps. 
The
solution then corresponds to the bound BEC under the joint action of
the harmonic and periodic
optical-lattice potentials.

\begin{figure}
\begin{center}
\includegraphics[width=1.\linewidth]{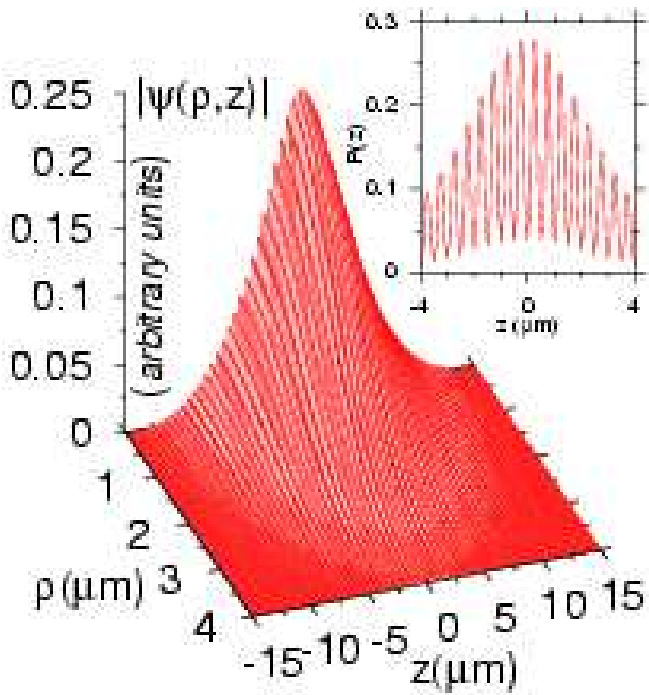}
\end{center}

\caption{The profile of the BEC wave function $|\psi(\rho,z)|$ formed in
the joint
optical-lattice and harmonic potentials for $n=- 0.3, s=4$, and 
$\alpha=1$.
In the off-set the one-dimensional probability $P(z)$ of this wave
function vs. $z$ is plotted for 4 $\mu$m  $>z>-4$ $\mu$m which clearly
exhibits the maxima 
and minima of the condensate wave function along the axial direction.}

\end{figure}

\section{Numerical Results} 

First we consider an attractive BEC formed in the combined harmonic and
periodic optical-lattice potentials for a specific attractive (negative)
nonlinearity $n$. Such a BEC is stable for $|n|$ less than a critical
value depending on the parameters of the optical-lattice potential
\cite{sad3}. We study the formation of a BEC in the combined harmonic and
optical-lattice potentials for $\alpha =1$ in Eq. (\ref{pot}) for a range
of values of $s$. Depending on the value of $\alpha$, the BEC could
be unstable for $|n|>0.4$ \cite{sad3} and for $\alpha =1$ we consider
$n=-0.3$. However,
for a larger $\alpha$, the BEC  is stable for a larger value of $|n|$
\cite{sad3} (see, for example, the stable condensate  studied in Fig. 5 
for $n=-0.5$ 
in the following 
for
$\alpha = 3$).  In Fig. 1 the plot of $|\psi(\rho,z)|$ vs. $\rho$ and $z$
illustrates the profile of the BEC bound state for $n=-0.3$ and
$s=4$. The BEC is in the form of narrow slices with the
optical-lattice barriers separating the slices. The large number of maxima
and minima due to the optical-lattice potential is not clearly visible in
this plot.  The maxima and minima in the axial direction are clear in the
plot of the one-dimensional probability $P(z)$ vs. $z$ in the off-set of
Fig. 1 for 4 $\mu$m $>z >-4$ $\mu$m. In this interval of $z$, there are 16
wells of the optical-lattice  potential and as many maxima and 
minima in $P(y,t)$.

\begin{figure}
 
\begin{center}
\includegraphics[width=.49\linewidth]{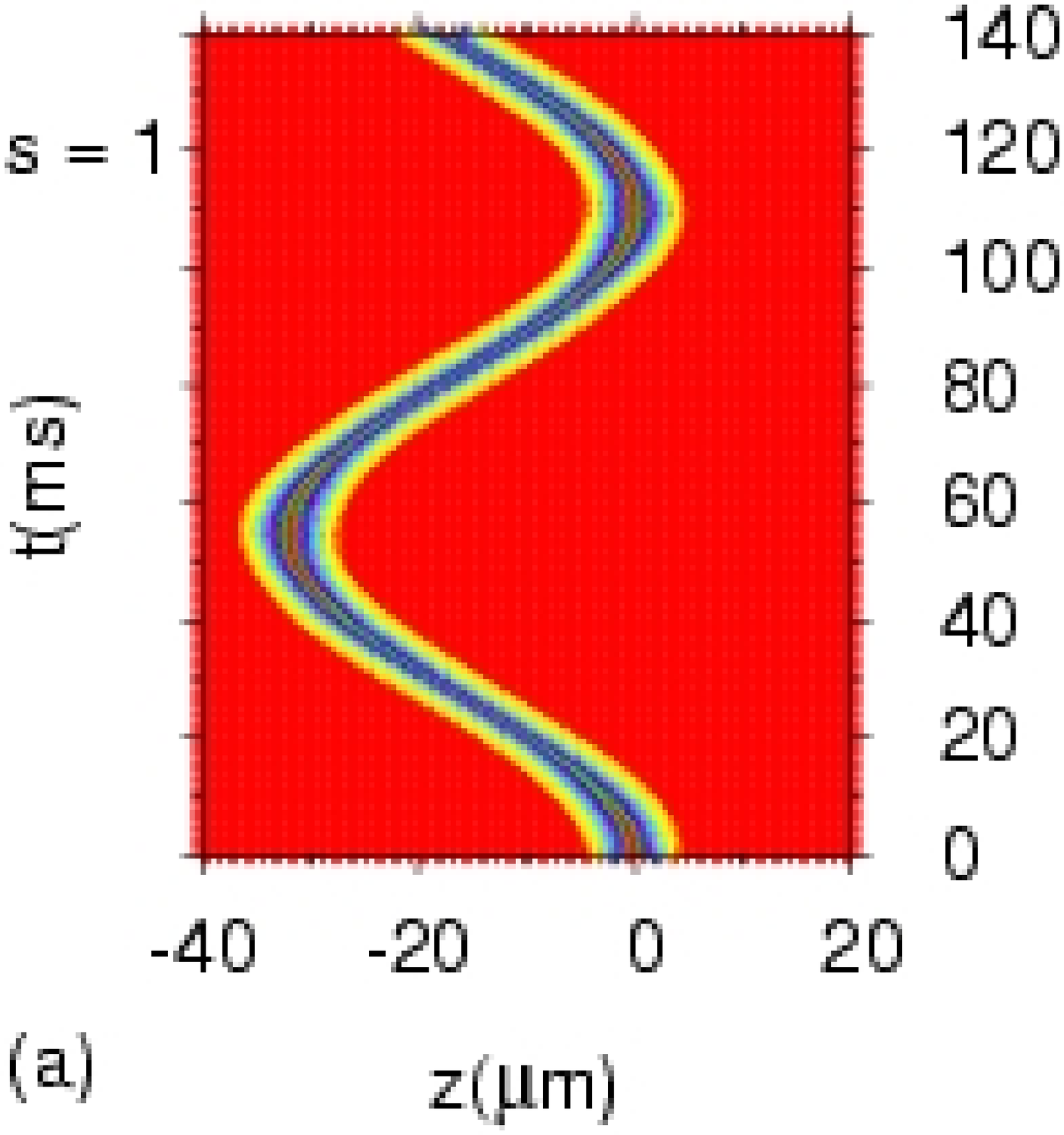}
\includegraphics[width=.49\linewidth]{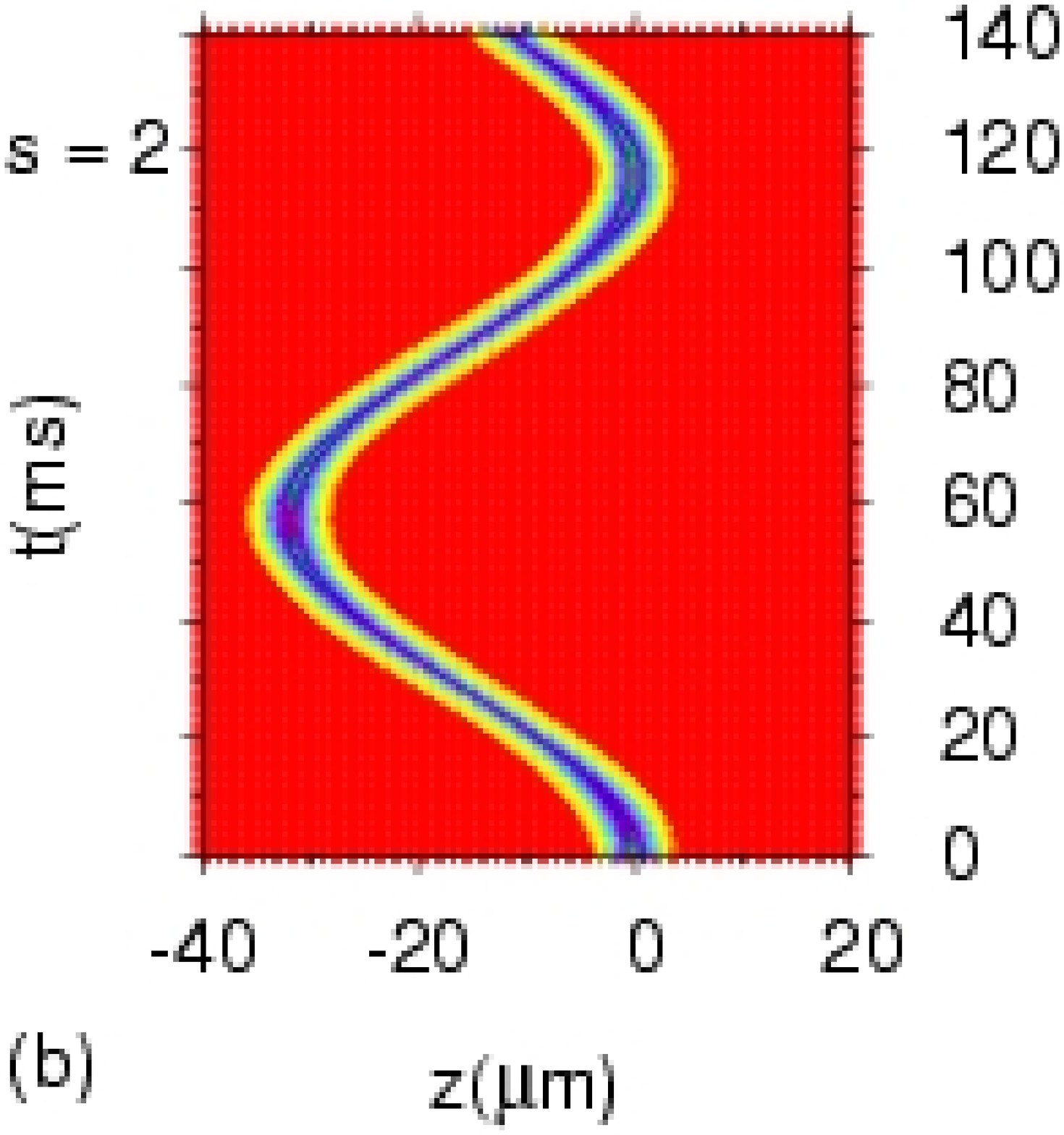}
\includegraphics[width=.49\linewidth]{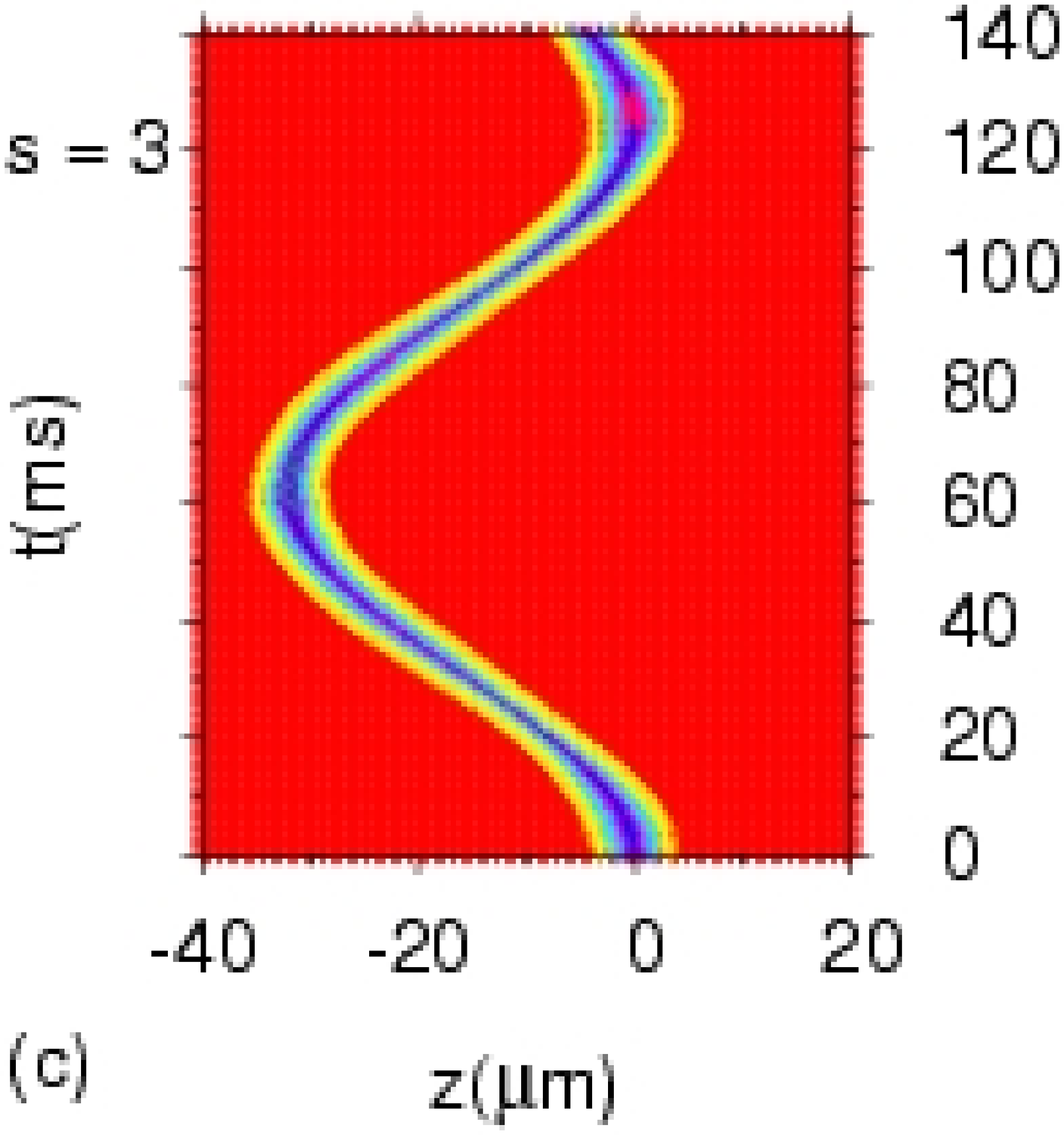}
\includegraphics[width=.49\linewidth]{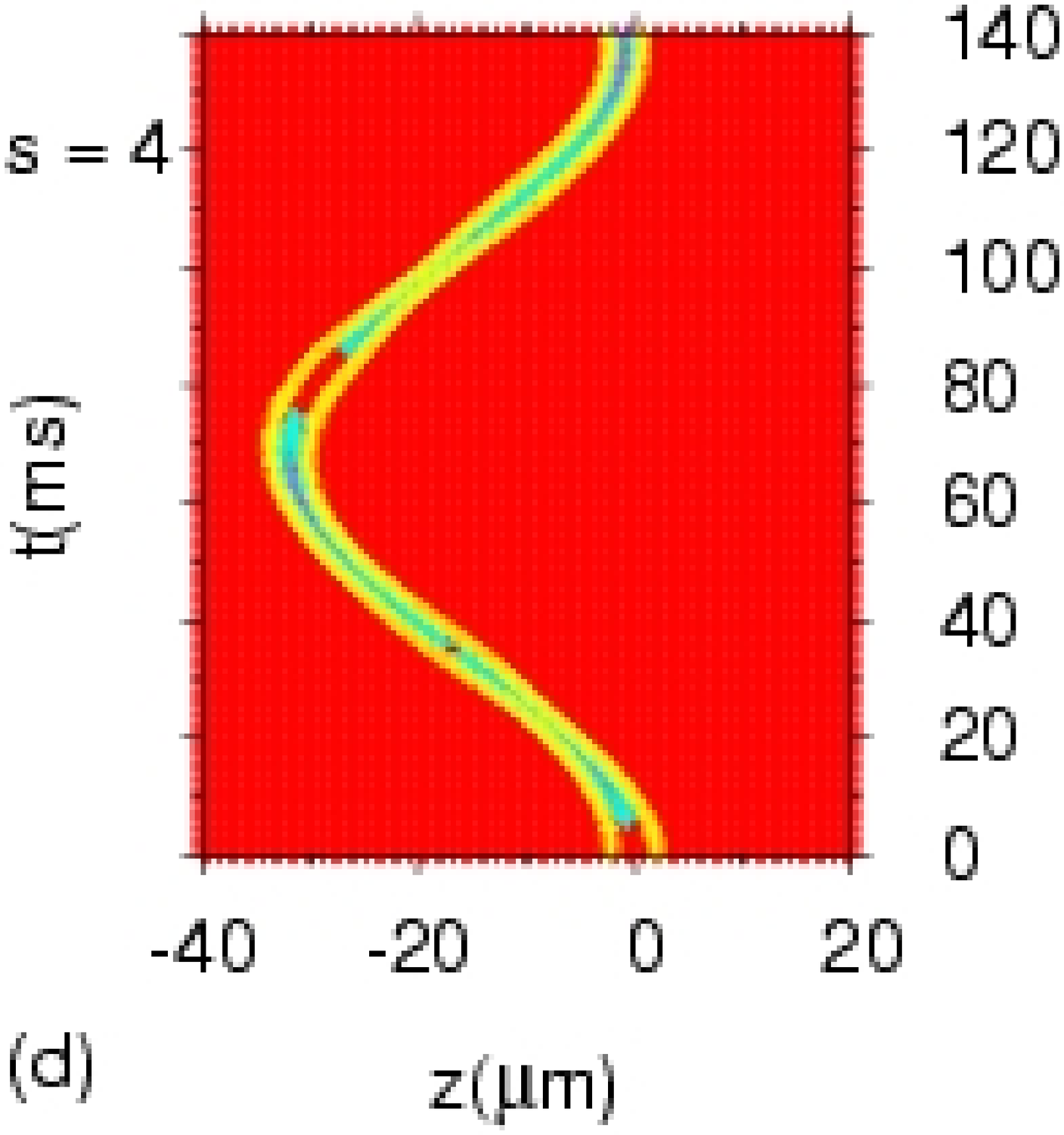}
\includegraphics[width=.49\linewidth]{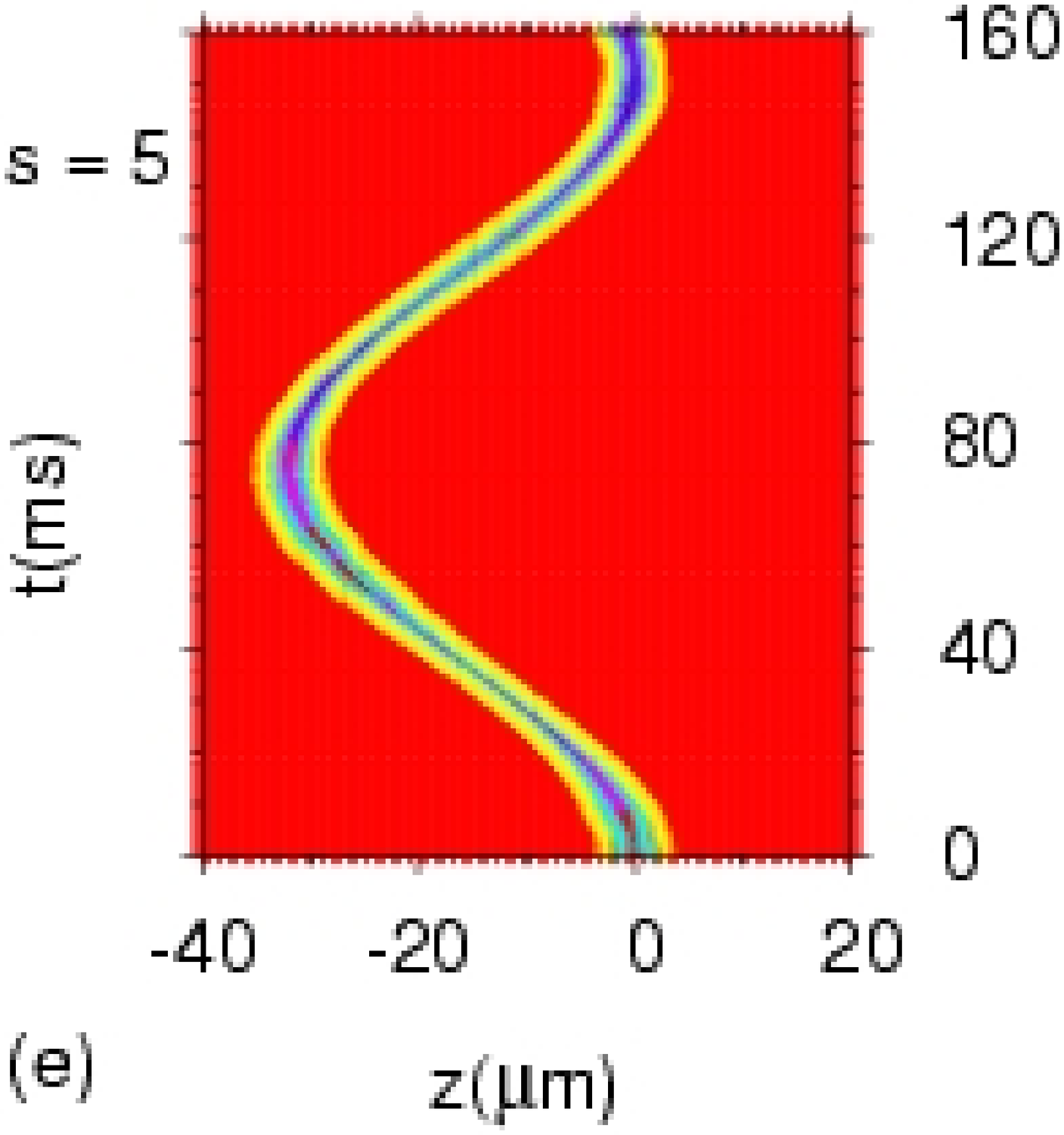}
\includegraphics[width=.49\linewidth]{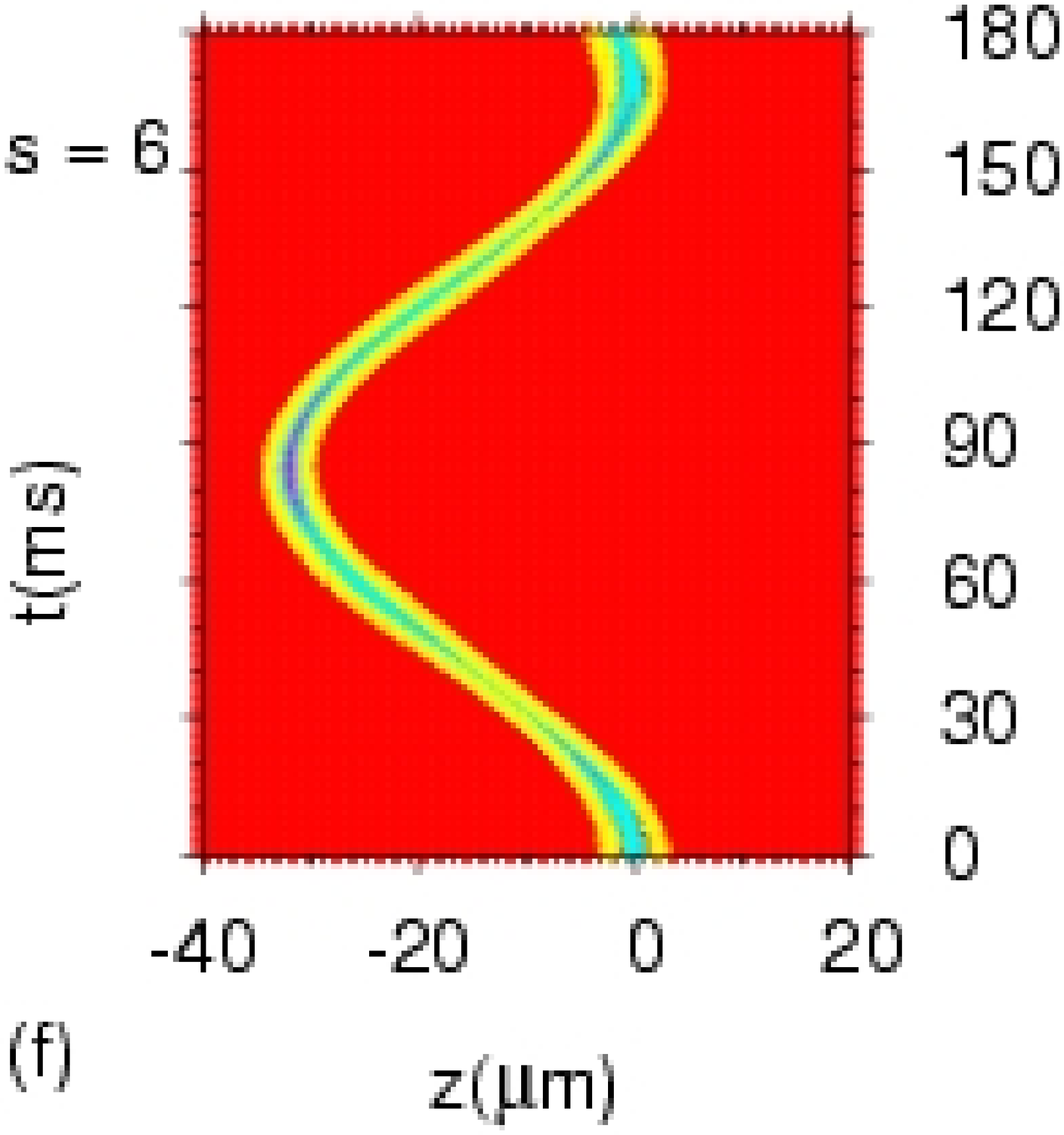}
\end{center}

\caption{(Color online) Contour plot of the one-dimensional probability
density $P(z,t)$ 
of the attractive BEC with $\alpha =1 $ and $n=-0.3$ executing a Josephson
oscillation when the
harmonic trap is suddenly displaced along the axial direction through a
distance 16 $\mu$m at $t=0$ for optical-lattice strength $s =$
(a) 1, (b) 2, (c) 3, (d) 4, (e) 5 and (f) 6. The period of Josephson
oscillation can be obtained from these plots. }
 
\end{figure}

Next we consider an oscillating BEC in the combined harmonic and
periodic
optical-lattice potentials. If
we suddenly displace the harmonic trap along the lattice axis by
a small distance after the formation of the BEC in the combined
potentials, the
condensate will acquire a potential energy, 
be out of equilibrium and start to oscillate. As the
height of the potential-well barriers of the optical-lattice potential is
much
larger than the energy of the system, the atoms in the condensate will
move by tunneling through the potential barriers. This fluctuating
transfer of  atoms across the potential barriers is due to Josephson
effect in a neutral quantum liquid.

\begin{figure}
 
\begin{center}
\includegraphics[width=1.\linewidth]{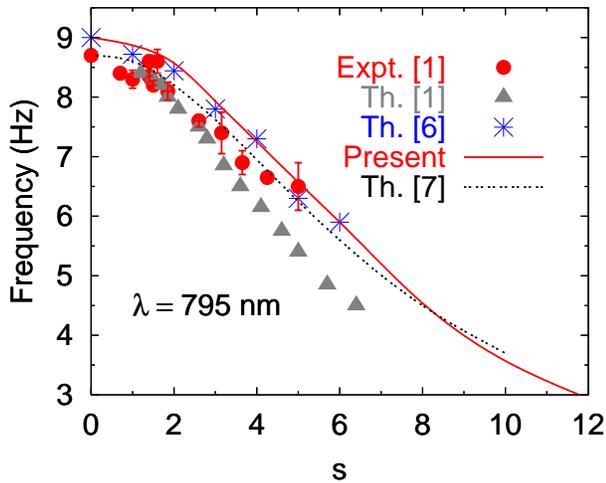}
\end{center}

\caption{(Color online) The frequency of the atomic current in the array
of Josephson
junctions as a function of optical-lattice strength $s$: $\bullet$
 with
error bar $-$ experiment of Cataliotti {\it et al.} \cite{cata} with
repulsive $^{87}$Rb atoms; 
$\triangle$    $-$ tight binding calculation taken
from Ref. \cite{cata}; $\star$
$-$ simulation with repulsive $^{87}$Rb atoms taken from
Ref. \cite{sad1};   
full line   - present calculation
with attractive atoms; 
dashed line - hydro-dynamical calculation with repulsive atoms taken from
Ref. \cite{str}.}

\end{figure}

With the attractive BEC of Fig. 1 we next study its Josephson 
oscillation
when the harmonic potential is suddenly displaced along the axial
direction 
by 16 $\mu$m. The BEC now acquires an added potential energy which it can
use to execute  a Josephson oscillation along the axial direction. The
Josephson oscillation is best studied numerically from the contour plot of
the one-dimensional  probability $P(z,t)$ vs $z$ and $t$ exhibited
in Figs. 2. These plots
clearly show the central position of the BEC along the axial $z$
direction. In the present simulation we take $\alpha =1$ and different
values of the optical-lattice strength $s$. These contour plots 
are very useful to find the Josephson frequencies. From
Figs. 2 the periods of Josephson oscillation are easily read off and the
frequencies calculated for different optical-lattice strengths $s$. 

In Fig. 3 we plot the Josephson frequencies vs. optical-lattice strength
$s$
from different studies for  $\alpha =1$. Specifically, in addition to the
present calculation with attractive atoms we also  show (1) the
experimental frequencies and corresponding tight-binding calculations for
repulsive atoms of
Cataliotti {\it et al.}   \cite{cata}, (2) the three-dimensional
simulation in the repulsive case from Ref. \cite{sad1}, and (3) the
hydro-dynamical calculation with repulsive atoms  from Ref. \cite{str}.

The most interesting conclusion from Fig. 3 is that the present
frequencies of
three-dimensional simulation for an attractive BEC are  practically the
same
as the frequencies of  the  repulsive BEC of Ref.
\cite{sad1}. Hence the Josephson frequencies are  independent of, or at
best weakly dependent on, 
the
number of atoms in the condensate and nature of interaction  (repulsive
or attractive). This is because in Eq. (\ref{d1}) the information about
the number of atoms and atomic interaction is contained solely in  the
nonlinearity parameter $n=Na/l$.  For $s=0$ the optical-lattice
potential is absent
and the condensate executes free oscillation with the frequency of the
axial potential (9 Hz) and the three-dimensional calculations by this
author
converges to this value in the $s \to 0$ limit for both
attractive and repulsive interactions. 

However, there are
some discrepancies between the present three-dimensional calculation and
the tight-binding calculation of Ref. \cite{cata} as well
as between the
present calculation and the hydro-dynamical calculation of Ref. \cite{str}
for  small values of optical-lattice strength $s$. 
The Josephson frequencies  of Refs. \cite{cata,str} do not seem to lead to
the same
result (9 Hz) as the present calculation in the low $s$ limit.  
The reason for this could be the  possible use of a 
slightly different axial trap frequency in these studies. 
For  large  values of optical-lattice strength $s$, there are no
experimental results. In this domain,  the 
present frequencies  agree well with the  hydro-dynamical calculation of
Ref. \cite{str}. But both these theoretical   results differ from the
tight-binding calculation of Ref. \cite{cata}. 
The reason for this  could be the use of a variational Gaussian 
ansatz  
in Ref. \cite{cata}.

Now we study the effect of the variation of laser wave
length on the Josephson frequency for a fixed strength
$s$ of the optical-lattice potential. To implement it we vary
the
parameter $\alpha$ of the optical-lattice potential (\ref{pot}), which
effectively 
varies the wave length  of the laser. 
The Josephson
frequency  is found to be weakly dependent on laser wave length  
for 1600 nm
$>\lambda > 600$ nm
leading to frequencies 
7.1 $\pm$ 0.3 Hz. If Josephson frequency were solely determined by the
effective strength of the optical-lattice potential $(4\pi^2/\lambda_0^2)
\times (s/\alpha^2)$ of Eq. (3), then from Fig. 3 a much larger variation
of frequency would be expected  for a variation of $\lambda$ in the range 
600 nm to 1600 nm. The  weak dependence  of Josephson frequency 
on $\alpha$ suggests that in addition to the strength of the
optical-lattice potential, the cosine term in Eq. (3) also plays a
decisive role in determining the Josephson frequency.

Next, we consider the effect of increasing the laser wave
length beyond  $\lambda=2000$ nm. 
The condensate in the periodic
optical-lattice  potential
is in the
form  of narrow slices with the optical-lattice barriers separating the
slices (see Fig. 1 and related description). These slices of the BEC have
to tunnel through the
optical-lattice barriers during Josephson oscillation. The increase in
laser wavelength corresponds to an increase in the widths of the
slices  of the BEC and optical-lattice barriers and to  execute a 
Josephson
oscillation, larger pieces of condensates have to tunnel freely  through
wider optical-lattice barriers.

With the increase of the laser wave length 
beyond a
critical value, 
the larger  pieces of condensates cannot freely tunnel
through wider barriers and there is actually a disruption of  
Josephson
oscillation. 
This is illustrated in Figs. 4 through a contour plot of
one-dimensional
probability density $P(z,t)$ with $\alpha =3$ and 4 
when the harmonic potential is suddenly displaced along the axial
direction through 16 $\mu$m at $t=0$ for $s=4$ for a small
nonlinearity $|n|=0.1$. The clean Josephson oscillation tracks seen in
Figs. 2 now disappear with the increase of laser wave length as can be
seen in Figs. 4 (a) and (b) where we show the contour  of the
one-dimensional
probability
density $P(z,t)$ for $\alpha =3$ and 4.  From Figs. 2 (d), 4 (a), and 4
(b) we see that for $s=4$ there is a clean Josephson
oscillation track for $\alpha =1$. The same is true for  $\alpha =2$.
However, with an
increase of  $\alpha $ to 3 the trail left by the condensate is not so
clean showing the beginning  of breakdown of Josephson oscillation in
Fig. 4 
(a). There is a complete breakdown (absence)  of Josephson oscillation
in
Fig. 4  (b) for $\alpha =4$. 

\begin{figure}
 
\begin{center}
\includegraphics[width=.9\linewidth]{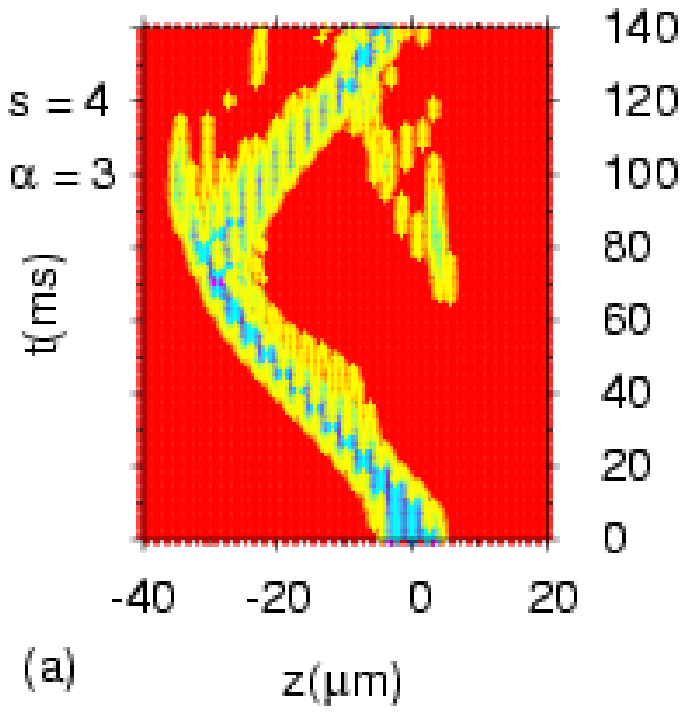}
\includegraphics[width=.9\linewidth]{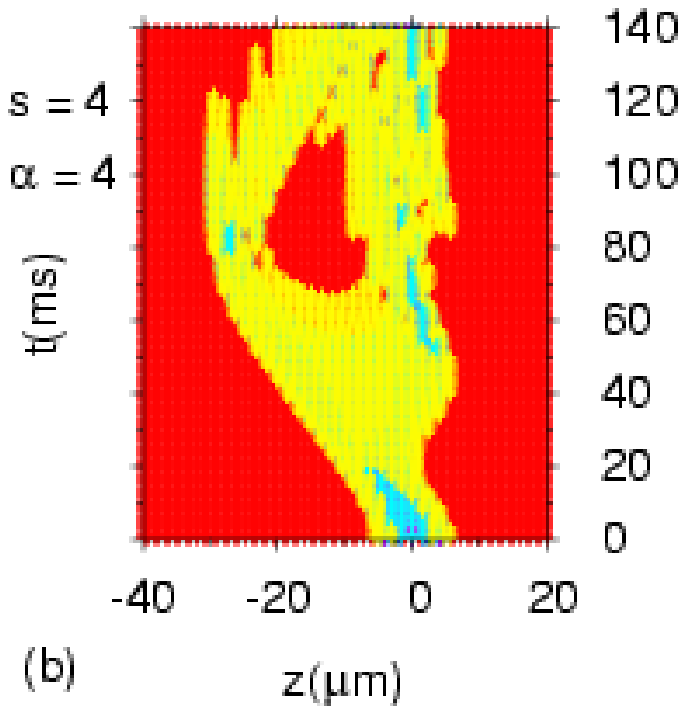}
\end{center}

\caption{(Color online) Contour plot of the one-dimensional probability
density $P(z,t)$
of the BEC with $s =4 $  and $n=-0.1$ when the
harmonic potential is suddenly displaced along the axial direction through
a
distance 16 $\mu$m at $t=0$ for laser wave length parameter  $\alpha =$
(a) 3 and (b) 4. The disruption of  Josephson
oscillation has already started for $\alpha =3$ and is complete for
$\alpha = 4$. }

\end{figure}

There has been a different account  of breakdown of Josephson
oscillation
in a neutral quantum fluid in Ref. \cite{bd} when the harmonic potential 
was
displaced by a very large distance ($>$ 130 $\mu$m) along the
periodic
optical-lattice  potential. It was suggested that the breakdown of
Josephson
oscillation    in  that study was due to a loss of phase coherence in the
condensate due to a classical transition from a superfluid  to an
``insulator" resulting in a modulational instability.
Other mechanisms for the loss of phase coherence in a condensate have also
been studied \cite{phco}. 
The loss of phase coherence considered  in all these investigations 
\cite{bd,phco} 
originated  from a  classical superfluid  to insulator transition,
different from a quantum transition of a superfluid to a Mott insulator
observed in Ref. \cite{mo}. 

From Fig. 3 we find that 
for a fixed laser wave length,  the Josephson frequency reduces
with increasing  optical-lattice strength, which seems plausible.
From Figs. 4 we find that for a fixed optical-lattice strength with an
increase in laser wave length 
the
Josephson oscillation is strongly attenuated. 
As the Josephson
oscillation takes place by quantum tunneling in atomic superfluid through
the optical-lattice barriers, this oscillation is bound to be reduced as
the height and width of the optical-lattice barriers are increased. This
reduction of Josephson oscillation with the increase of the parameters
$s$ or $\alpha$  will first manifest in a reduction  of  
Josephson
frequency and eventually to a disappearance of Josephson oscillation. 
This reduction in frequency is explicit in Fig. 3. A breakdown of
Josephson oscillation with increasing $\alpha$ is found in Fig. 4. 
The  possible breakdown of Josephson oscillation with large $s$
for $\alpha =1$ is not investigated in this paper.

Finally,  we investigate if the above disruption of Josephson
oscillation
for a large $\alpha $ could provoke a collapse in an attractive BEC. We
find in our numerical study that in the cases where there is a clean
Josephson oscillation there is
no collapse in an attractive condensate. 
However, for large $\alpha$,
when there is a disruption of Josephson oscillation, e.g., as in
Figs. 4,
one could have a collapse  in an attractive condensate with sufficiently
strong nonlinearity. 

\begin{figure}
 
\begin{center}
\includegraphics[width=.85\linewidth]{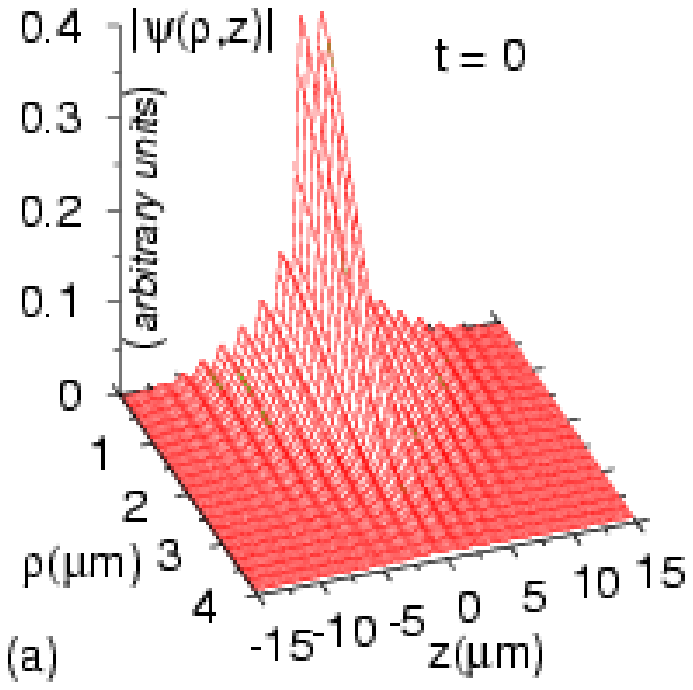}
\includegraphics[width=.85\linewidth]{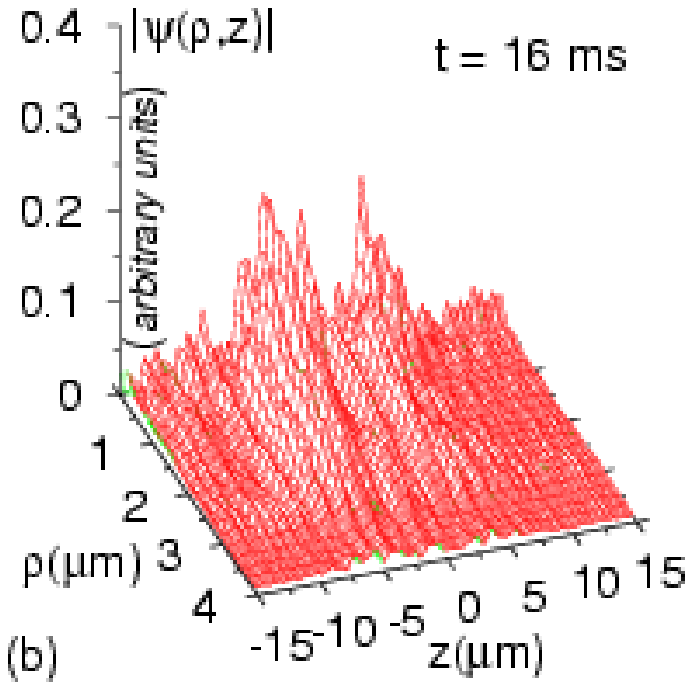}
\end{center}

\caption{(a) The profile of the wave function $|\psi(\rho,y)|$ formed in
the joint optical-lattice  and harmonic potentials 
for $s
=4$, $\alpha =3$, and $n=-0.5$ at $t=0$. (b) The profile of the wave
function of (a) at $t = 16$ $\mu$m after an axial displacement of the
harmonic potential through 16 $\mu$m at $t=0$. }

\end{figure}

To demonstrate the above  claim we performed the simulation
of Josephson oscillation using the parameters used in    Figs. 2 for 
attractive nonlinearity close to the critical nonlinearity
for collapse \cite{sad3}. In the actual conditions of  Figs. 2, 
the  Josephson oscillation 
is  always perfect and there is no
collapse in the condensate provoked by  this oscillation. 
However, the scenario changes as the parameter $\alpha $ is
increased for a large nonlinearity $|n|$. To illustrate this we performed
calculation with
optical-lattice strength $s =4$, $\alpha =3$, and nonlinearity 
$n=-0.5$. The stationary BEC wave function in this case is localized
and 
stable as shown in Fig. 5 (a) at $t=0$. A Josephson oscillation is then
initiated by displacing the harmonic potential  along the axial direction
through a distance of 16 $\mu$m. Soon after this displacement, due to the
onset of collapse  
there is no stationary solution to the GP equation. Numerically, 
the nonexistence of a stationary solution to the GP equation and the
subsequent  nonequilibrium dynamics of the condensate 
manifests in a destruction (delocalization) of the localized
stationary wave function at $t=0$ and a generation of  a dynamical
nonequilibrium wave function spread
over the entire spatial domain of
discretization as seen in Fig. 5 (b) at $t=16$ ms.

When there is a free Josephson tunneling, the condensate 
oscillates as a whole without significant modification of shape or local
density. However, when there is a disruption of 
Josephson tunneling and the BEC is given additional potential 
energy to
oscillate by displacing the harmonic potential, the condensate is
locally squeezed to a
smaller region in space resulting in a very large density and thus
inducing a collapse of the condensate as illustrated in Figs. 5.

The investigations of Refs. \cite{str,str1} also provides answer to some
of
the questions raised in the present paper, e.g., how does the Josephson
oscillation depend on the nonlinearity parameter $n$ and the
optical-lattice 
potential strength $s$. Using a hydro-dynamical model, the authors of Ref.
\cite{str} show for an elongated or cigar-shaped confinement ($\nu<<1$)
that the frequency of oscillation $\omega_z\equiv \nu \omega$ in a pure
harmonic potential gets changed in the presence of the optical-lattice
potential according to $\omega_{\mbox{opt}} = \omega_z\sqrt{m/m^*}$, where
$m$ is the atomic mass and $m^* $ is an effective mass.  A good estimate
of $m/m^*$ can be made by neglecting both the magnetic trapping and the
atom-atom interaction from the spectrum of the linear Schr\"odinger
equation for the one-dimensional periodic optical-lattice  potential. In
this
approximation $m/m^*$ is a universal function of the parameter $s$,
calculated in Ref. \cite{str}. 
This dependence of $m/m^*$ on $s$ has been
used to calculate the dependence of Josephson frequency on $s$ \cite{str}
as quoted in Fig. 3 of this paper. 
 The quantity $m/m^*$ also depends on
nonlinearity $n$ and harmonic oscillator frequency $\omega$. However, the
dependence of $m/m^*$ on $n$ and $\omega$ is weak for small values of
these parameters \cite{str1}.  
In Ref. \cite{str1} the authors study
the dependence of $m/m^*$ on nonlinearity $n$ in a model
and demonstrate the weak dependence for small and medium values of $n$.
Consequently, the independence of the Josephson frequency on the
nonlinearity parameter $n$ as well as the dependence of this frequency on
the parameter $s$ 
as noted in Fig. 3 
seems to be consistent with Refs.
\cite{str,str1}. However, it should be noted that the studies of Refs.
\cite{str,str1} did not consider Josephson oscillation for attractive
interaction and the possibility of collapse as in this paper. Hence the
conclusions  of  Refs. \cite{str,str1} are only valid for repulsive
atomic interaction and the present study extends them to the case
of
attractive
atomic interactions.
Moreover,
the present study is based on the three-dimensional
mean-field microscopic GP equations, whereas those of
Refs. \cite{str,str1}
are based on  hydro-dynamical  
and  one-dimensional mean-field
models 
of the condensate.

\section{Summary and Conclusion}

We have performed numerical simulation based on an axially-symmetric
mean-field GP
equation to address different aspects of Josephson oscillation of an
attractive  BEC in
a combined harmonic and periodic
optical-lattice  potentials. The latter potential is
created by a standing-wave laser beam.
We find that the
Josephson frequency  is virtually independent of the number of
atoms
in the condensate and the nature of atomic interaction: attractive or
repulsive. The  mean-field results for Josephson frequencies for
different optical-lattice strengths are in qualitative  agreement with
experiment
\cite{cata} and other calculations \cite{cata,sad1,str}. However,
there is some quantitative disagreement between different results. 
The
Josephson oscillation continues with essentially unchanged frequency 
($7.1\pm 0.3$ Hz) for a variation of laser wave length in the region  
1600 nm $>\lambda >$ 600 nm. 
However, as $\lambda$ is increased past 2000
nm  there is a
disruption of Josephson oscillation. For an attractive BEC with
nonlinearity stronger than a critical value 
this disruption
of Josephson oscillation could induce a collapse. These features of
Josephson oscillation could be tested experimentally and will provide a
test for the mean-field model.

\vskip 1cm
\acknowledgments

{The work was supported in part by the CNPq 
of Brazil.}


 \end{document}